\documentclass[prl,twocolumn,showpacs,floatfix,amsfonts]{revtex4}
\usepackage{graphicx,graphics,color,epsfig}% Include figure files
\usepackage{bm}
\usepackage{amsmath}
\usepackage{amssymb}
\begin{document}
\preprint{}
\title{Josephson Current in the Presence of  a  Precessing Spin}
\author{Jian-Xin Zhu}
\affiliation{Theoretical Division, Los Alamos National Laboratory,
Los Alamos, New Mexico 87545}
\author{A. V. Balatsky}
\affiliation{Theoretical Division, Los Alamos National Laboratory,
Los Alamos, New Mexico 87545}

\begin{abstract}
The Josephson current in the presence of a precessing spin between
various types of superconductors is studied. It is shown that the
Josephson current flowing between two spin-singlet pairing
superconductors is not modulated by the precession of the spin.
When both superconductors have equal-spin-triplet pairing state,
the flowing Josephson current is modulated with twice of the
Larmor frequency by the precessing spin. It was also found that up
to the second tunneling matrix elements, no Josephson current can
occur with only a  direct exchange interaction between the
localized spin and the conduction electrons, if the two
superconductors have different spin-parity pairing states.
\end{abstract}
\maketitle

There is a growing interest in a number of techniques that allow
one to detect and manipulate a single spin in the solid state.
Partial list includes optical detection of ESR in a single
molecule~\cite{Koehler93}, tunneling through a quantum 
dot~\cite{Engel01}, and, more recently, electron-spin-resonance-scanning
tunneling microscopy (ESR-STM)
technique~\cite{Mana89,Mana00,Durkan02,Manoharan02}. There is a
growing recognition that the ESR-STM technique  is capable of
detecting the precession of a single spin through the modulation
of the tunnel current.  Interest in ESR-STM lies in the
possibility to detect and manipulate a single
spin~\cite{Manoharan02}, which is crucial in spintronics and
quantum information processing.

Several proposals have been made for the mechanism of the spin
detection with the ESR-STM.  One is the effective spin-orbit
interaction of the conduction electrons in the two-dimensional
surface coupling the injected unpolarized current to the
precessing spin~\cite{Balatsky02a}. Another one is the
interference between two resonant tunneling components through the
magnetic field split Zeeman levels~\cite{Mozy01}. Both of these
mechanisms rely on a spin-orbit coupling to couple the local spin
to the conduction electrons and have assumed no spin polarization
of tunneling electrons. On the other hand, one can perform ESR-STM
measurements on samples with much 
smaller spin-orbit
coupling~\cite{Durkan02}. Theoretically it is also important to
investigate the role of direct exchange  in ESR-STM measurements, without
any
spin-orbit coupling~\cite{Balatsky02b}.  Exchange interaction has a
tremendous
effects on the physics of conducting substances when magnetic
impurities are present~\cite{Li98,Manoharan00}.

The above mentioned experimental and theoretical studies are
concentrated on the tunneling between two normal metals. A natural
extension  is a  question   of  the role of a precessing spin
localized inside a tunneling barrier on the Josephson current
between two weakly coupled superconductors. This is the problem we
address in this Letter.

Previously, the Josephson effect between superconductors with
nontrivial pairing symmetry has been extensively studied, see
e.g.,~\cite{Sigrist91}. There are two main aspects
of current study that differ from the previous work: 1) we will
consider the effect of the precessing localized spin in the
junction on the Josephson current. This effect, to our knowledge,
has not been addressed before; 2) we will assume {\em no 
spin-orbit coupling} between the superconductors. The role of the
spin-orbit coupling will be addressed elsewhere.

\begin{figure}
\centerline{\psfig{figure=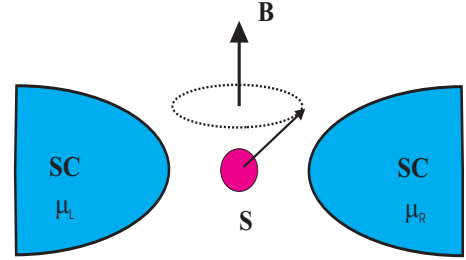,height=3.5cm,width=6cm,angle=0}}
\caption{Magnetic spin coupled to two superconducting leads. In
the presence of a magnetic field ${\bf B}$, the spin precesses
around the field direction.} \label{FIG:SETUP}
\end{figure}

The model system under consideration is illustrated in
Fig.~\ref{FIG:SETUP}. It consists of two ideal superconducting
leads coupled to each other by a single magnetic spin. In the
presence of a magnetic field, the spin precesses around the field
direction.  We  neglect the interaction of the spin with two
superconducting leads.  The Hamiltonian for the
Josephson junction can then be generally written
as~\cite{Sigrist91}:
\begin{equation}
H=H_{L}+H_{R}+H_{T}\;.
\end{equation}
The first two terms are respectively the Hamiltonian for electrons
in the left and right superconducting leads of the tunneling
junction:
\begin{eqnarray}
&H_{L(R)}=\sum_{\mathbf{k}(\mathbf{p});\sigma}\epsilon_{\mathbf{k}(\mathbf{p})}
c_{\mathbf{k}(\mathbf{p}),\sigma}^{\dagger}c_{\mathbf{k}(\mathbf{p}),\sigma}&
 \nonumber \\
& +\frac{1}{2}\sum_{\mathbf{k}(\mathbf{p});\sigma,\sigma^{\prime}}
[\Delta_{\sigma\sigma^{\prime}}(\mathbf{k}(\mathbf{p}))
c_{\mathbf{k}(\mathbf{p}),\sigma}^{\dagger}
c_{-\mathbf{k}(-\mathbf{p}),\sigma^{\prime}}^{\dagger}
+\mbox{H.c.}]\;,&
\end{eqnarray}
where we have denoted the electron creation (annihilation)
operators in the left (L) lead by $c_{\mathbf{k}\sigma}^{\dagger}$
($c_{\mathbf{k}\sigma}$) while those in the right (R) lead by
$c_{\mathbf{p}\sigma}^{\dagger}$ ($c_{\mathbf{p}\sigma}$). The
quantities $\mathbf{k}$ ($\mathbf{p})$ are momenta and $\sigma$ is
the spin index. The quantities
$\epsilon_{\mathbf{k}(\mathbf{p}),\sigma}$,
$\Delta_{\sigma\sigma^{\prime}}(\mathbf{k}(\mathbf{p}))$ are,
respectively, the single particle energies of conduction
electrons, and the pair potential (also called gap function) in
the leads. For the purpose of this work, the physical origin for
the superconducting instability is beyond the scope of our
discussion. The two leads are weakly coupled with the tunneling
Hamiltonian:
\begin{equation}
H_{T}=\sum_{\mathbf{k},\mathbf{p};\sigma,\sigma^{\prime}}
[T_{\sigma\sigma^{\prime}}(\mathbf{k},\mathbf{p})
c_{\mathbf{k}\sigma}^{\dagger}c_{\mathbf{p}\sigma^{\prime}}
+\mbox{H.c.}]\;,
\end{equation}
where the tunneling matrix element
$T_{\sigma\sigma^{\prime}}(\mathbf{k},\mathbf{p})$ transfer
electrons through an insulating barrier. When a local spin is
embedded into the tunneling barrier, the tunneling matrix can be
written in the spin space as~\cite{Balatsky02b}:
\begin{equation}
\hat{T}=T_0 \exp\left[ -\sqrt{\frac{\Phi-J\mathbf{S}\cdot
\hat{\bm{\sigma}}}{\Phi_0}}\right]\;,
\end{equation}
where $\Phi$ is the spin-independent potential barrier, and
$\Phi_0=\hbar^{2}/2m_e d^{2}$ is the characteristic energy scale for
the barrier width $d$, $J$ is the exchange interaction between the
local spin $\mathbf{S}$ and the tunneling electrons denoted by the
Pauli matrix $\hat{\bm{\sigma}}$. In an external magnetic field
$\mathbf{B}$, a torque will act on the magnetic moment $\bm{\mu}$
of amount $\bm{\mu}\times \mathbf{B}$, where
$\bm{\mu}=\gamma\bm{S}$ with $\gamma$ the gyromagnetic ratio. The
equation of motion of the local spin is given by
$\frac{d\bm{\mu}}{dt}=\bm{\mu} \times (\gamma \mathbf{B})$. For a
static magnetic field applied along the $z$ direction, we shall
see that the local spin would precess about the field at the
Larmor frequency $\omega_{L}=\gamma B$, i.e.,
$\mathbf{S}=\mathbf{n}(t)S$, where $S$ is the magnitude of the
local spin and $\mathbf{ n}(t)=(n_x,n_y,n_z)=(n_{\perp}
\cos(\omega_{L}t),-n_{\perp}\sin(\omega_{L}t),n_{\parallel})$ the
unit vector for the `instantaneous' spin orientation. Here
$n_{\parallel}$ and $n_{\perp}$ are the magnitude of the
longitudinal and transverse components of $\mathbf{S}$ to the
field direction. They obey the sum rule
$n_{\parallel}^{2}+n_{\perp}^{2}=1$. We note that the expression
for $\mathbf{n}(t)$ shows the constant left-handed precession, and
the $z$ component of $\mathbf{S}$ is time-independent. The
precession of the spin can also be obtained quantum mechanically
by replacing the local spin operator with its average value. The
exchange term in the exponent of the tunneling matrix element is
very small as compared with the barrier height $\Phi$. We then
perform the Taylor expansion in $JS$ and arrive at:
\begin{eqnarray}
\hat{T}&=&T_0 \exp\left( -\sqrt{\frac{\Phi}{\Phi_0}}\right)
\left[\cosh \bigglb{(}
\frac{JS}{2\Phi}\sqrt{\frac{\Phi}{\Phi_0}}\biggrb{)}\right.
\nonumber
\\ &&\left. +\mathbf{n}(t)\cdot
\hat{\bm{\sigma}}\sinh\bigglb{(}
\frac{JS}{2\Phi}\sqrt{\frac{\Phi}{\Phi_0}}\biggrb{)}\right]\;.
\label{EQ:SPIN}
\end{eqnarray}

Since the energy associated with the spin precession,
$\hbar\omega_{L}\sim 10^{-6}\;\mbox{eV}$ is much smaller than the
typical electronic energy on the order of 1 eV, the spin
precession is very slow as compared to the time scale of all
conduction electron process. This fact allows us to treat the
electronic problem adiabatically as if the local spin is static
for every instantaneous spin orientation~\cite{Comment1}. Our
remaining task is to calculate the Josephson current in the presence of 
the spin. 
%In the first order perturbation in the tunneling matrix
%element $T_{\sigma\sigma{\prime}}$, the current can be calculated
%as:
%\begin{equation}
%I(t)=-i\int_{-\infty}^{t}\langle[\hat{I}_{H}(t),H_{TH}(t^{\prime})]_{-}
%\rangle dt^{\prime}\;.
%\end{equation}
%where the brackets $[\dots]_{-}$ stands for a commutator and
%$\langle \dots \rangle$ for the thermal average over the states of
%the unperturbed Hamiltonian of the system $H_{0}=H_{L}+H_{R}$. The
%operator inside the commutator is defined in the Heisenberg
%picture as
%$\mathcal{O}_{H}(t)=e^{iH_{0}t}\mathcal{O}e^{-iH_{0}t}$. 
The
current operator is given by
\begin{equation}
\hat{I}=ie\sum_{\mathbf{k},\mathbf{p};\sigma,\sigma^{\prime}}
[T_{\sigma\sigma^{\prime}}(\mathbf{k},\mathbf{p};t)
c_{\mathbf{k}\sigma}^{\dagger}
c_{\mathbf{p}\sigma^{\prime}}-\mbox{H.c.}]\;.
\end{equation}
When a voltage bias $eV=\mu_{L}-\mu_{R}$ is applied across the junction, 
%However, the
%energy in the initial Hamiltonian is measured on an absolute
%energy scale.
%It would be more convenient to the calculation for the energy to
%be measured relative to the respective chemical potential on each
%side of the junction. For this purpose, a new Hamiltonian
%$K_{L(R)}=H_{L(R)}-\mu_{L(R)}N_{L(R)}$ with
%$N_{L(R)}=\sum_{\mathbf{k}(\mathbf{p}),\sigma}
%c_{\mathbf{k}(\mathbf{p})\sigma}^{\dagger}
%c_{\mathbf{k}(\mathbf{p})\sigma}$ is introduced. 
following the standard procedure~\cite{Ambegaokar63}, we can write
the phase dependent contribution, i.e., the Josephson current as:
\begin{eqnarray}
I_{J}(t)&=&e\int_{-\infty}^{t} dt^{\prime} [e^{ieV(t+t^{\prime})}
\langle [A(t),A(t^{\prime})]_{-}\rangle \nonumber \\
&& -e^{-ieV(t+t^{\prime})} \langle
[A^{\dagger}(t),A^{\dagger}(t^{\prime})]_{-}\rangle]\;,
\label{EQ:JOSEPHSON}
\end{eqnarray}
where the operator
$A(t)=\sum_{\mathbf{k},\mathbf{p};\sigma,\sigma^{\prime}}
T_{\sigma\sigma^{\prime}}(\mathbf{k},\mathbf{p};t)
\tilde{c}_{\mathbf{k}\sigma}^{\dagger}(t)
\tilde{c}_{\mathbf{p}\sigma^{\prime}}(t)$.
%\begin{equation}
%A(t)=\sum_{\mathbf{k},\mathbf{p};\sigma,\sigma^{\prime}}
%T_{\sigma\sigma^{\prime}}(\mathbf{k},\mathbf{p};t)
%\tilde{c}_{\mathbf{k}\sigma}^{\dagger}(t)
%\tilde{c}_{\mathbf{p}\sigma^{\prime}}(t)\;.
%\end{equation}
Here the operators $\tilde{c}_{\mathbf{k}(\mathbf{p})\sigma}(t)
=e^{iK_{L(R)}t} c_{\mathbf{k}(\mathbf{p})\sigma}e^{-iK_{L(R)}t}$ 
with $K_{L(R)}=H_{L(R)}-\mu_{L(R)}N_{L(R)}$ and  
$N_{L(R)}=\sum_{\mathbf{k}(\mathbf{p}),\sigma}
c_{\mathbf{k}(\mathbf{p})\sigma}^{\dagger}
c_{\mathbf{k}(\mathbf{p})\sigma}$.

For either spin-singlet or spin-triplet superconductors, we can
perform the Bogoliubov transformation to express the electron
operators in terms of quasiparticle operators:
\begin{equation}
c_{\mathbf{k}\sigma}=\sum_{\sigma^{\prime}}
(u_{\mathbf{k}\sigma\sigma^{\prime}}\gamma_{\mathbf{k}\sigma^{\prime}}
-\sigma v_{-\mathbf{k}\sigma\sigma^{\prime}}^{*}
\gamma_{-\mathbf{k}\sigma^{\prime}}^{\dagger})\;
\end{equation}
to diagonalize the unperturbed Hamiltonian, where
$(u_{\mathbf{k}\sigma\sigma^{\prime}},
v_{\mathbf{k}\sigma\sigma^{\prime}})^{T}$  is the Bogoliubov
quasiparticle wavefunction. For a spin singlet superconductor, the
order parameter matrix can be written as:
%\begin{equation}
$\hat{\Delta}(\mathbf{k})=(i\hat{\sigma}_{y})\psi(\mathbf{k})$,
%=\left(\begin{array}{cc} 0 & \psi(\mathbf{k}) \\
%-\psi(\mathbf{k}) & 0
%\end{array}\right)\;,
%\end{equation}
where $\psi(\mathbf{k})$ is an even function of $\mathbf{k}$. The
quasiparticle wavefunction is then given by:
\begin{equation}
\left( \begin{array}{c} u_{\mathbf{k}\sigma\sigma^{\prime}}\\
v_{\mathbf{k}\sigma\sigma^{\prime}} \end{array} \right) =\left(
\begin{array}{c}
u_{\mathbf{k}}
e^{i(\varphi_{\mathbf{k}}+\varphi)}\delta_{\sigma\sigma^{\prime}}
\\
v_{\mathbf{k}} \delta_{\sigma,-\sigma^{\prime}} \end{array}
\right)\;, \label{Eq:UV}
\end{equation}
with
\begin{equation}
\left( \begin{array}{c} u_{\mathbf{k}} \\ v_{\mathbf{k}}
\end{array} \right)
=\left( \begin{array}{c}
 \sqrt{\frac{1}{2}\left(1+
\frac{\xi_{\mathbf{k}}}{E_{\mathbf{k}}}\right)} \\
\sqrt{\frac{1}{2}\left(1-
\frac{\xi_{\mathbf{k}}}{E_{\mathbf{k}}}\right)}
\end{array} \right)\;,
\label{EQ:UV-0}
\end{equation}
where we have introduced $\psi(\mathbf{k})=\vert
\psi(\mathbf{k})\vert e^{i(\varphi_{\mathbf{k}}+\varphi)}$ with
$\varphi_{\mathbf{k}}$ and $\varphi$ being the internal and global
phase, and $\xi_{\mathbf{k}}=\epsilon_{\mathbf{k}}-\mu$, and
$E_{\mathbf{k}}=\sqrt{\xi_{\mathbf{k}}^{2}+\vert
\psi(\mathbf{k})\vert^{2}}$. For the spin-triplet pairing state,
the order parameter can be written as:
%\begin{eqnarray}
$\hat{\Delta}(\mathbf{k})=i(\mathbf{d}(\mathbf{k})\cdot
\hat{\bm{\sigma}})\hat{\sigma}_{y} $,
%\nonumber \\
%& =&\left(
%\begin{array}{cc}
%-d_{u}(\mathbf{k})+id_{v}(\mathbf{k}) & d_{w}(\mathbf{k}) \\
%d_{w}(\mathbf{k}) & d_{u}(\mathbf{k})+id_{v}(\mathbf{k})
%\end{array} \right)\;,
%\end{eqnarray}
where $\mathbf{d}=(d_u,d_v,d_w)$ is an odd vectorial function of
$\mathbf{k}$ defined in a three-dimensional spin space spanned by
$(u,v,w)$. We shall be typically concerned with two types of
triplet pairing states---non-equal-spin pairing, where the Cooper
pairs are formed by electrons with anti-parallel spins, and
equal-spin pairing, where the Cooper pairs are formed by electrons
with parallel spins. The non-equal spin pairing state has the
form:
\begin{equation}
\hat{\Delta}(\mathbf{k})=\left(
\begin{array}{cc}
0 & d_{I}(\mathbf{k})\\
d_{I}(\mathbf{k}) & 0
\end{array}
\right)\;,
\end{equation}
corresponding to $(d_{u},d_{v},d_{w})=(0,0,d_{I}(\mathbf{k}))$.
This type of pairing state may be realized in the recently
discovered superconducting
Sr$_2$RuO$_4$~\cite{Maeno01,Agterberg97}. The equal spin pairing
state has the form:
\begin{equation}
\hat{\Delta}(\mathbf{k})=\left(
\begin{array}{cc}
2d_{II}(\mathbf{k}) & 0 \\
0 & 2d_{II}(\mathbf{k})
\end{array}
\right)\;,
\end{equation}
corresponding to
$(d_{u},d_{v},d_{w})=(0,-i2d_{II}(\mathbf{k}),0)$. This state may
be relevant to the A-phase of superfluid
$^{3}$He~\cite{Anderson61} and of heavy fermion
UPt$_{3}$~\cite{Joynt02}. A little algebra yields the
quasiparticle wavefunction:
\begin{equation}
\left( \begin{array}{c} u_{\mathbf{k}\sigma\sigma^{\prime}}\\
v_{\mathbf{k}\sigma\sigma^{\prime}} \end{array} \right) =\left(
\begin{array}{c}
\sigma u_{I,\mathbf{k}}
e^{i(\varphi_{\mathbf{k}}+\varphi)}\delta_{\sigma\sigma^{\prime}}
\\
v_{I,\mathbf{k}} \delta_{\sigma,-\sigma^{\prime}} \end{array}
\right)\;, \label{EQ:UV-I}
\end{equation}
for the non-equal-spin-triplet pairing state; while
\begin{equation}
\left( \begin{array}{c} u_{\mathbf{k}\sigma\sigma^{\prime}}\\
v_{\mathbf{k}\sigma\sigma^{\prime}} \end{array} \right) =\left(
\begin{array}{c}
\sigma u_{II,\mathbf{k}}
e^{i(\varphi_{\mathbf{k}}+\varphi+\pi)}\delta_{\sigma\sigma^{\prime}}
\\
v_{II,\mathbf{k}} \delta_{\sigma\sigma^{\prime}} \end{array}
\right)\;, \label{EQ:UV-II}
\end{equation}
for the equal-spin-triplet pairing state. Here
$(u_{I(II),\mathbf{k}},v_{I(II),\mathbf{k}})^{T}$ has the same
form as that given by Eq.(\ref{EQ:UV-0}) except
$E_{\mathbf{k}}=\sqrt{\xi_{\mathbf{k}}^{2}+\vert
d_{I,II}(\mathbf{k})\vert^{2}}$ and $d_{I,II}(\mathbf{k})=\vert
d_{I,II}(\mathbf{k})\vert e^{i(\varphi_{\mathbf{k}}+\varphi)}$,
respectively. Due to the opposite parity of the triplet pairing
state as compared with the singlet counterpart, there appears an
additional factor $\sigma$ ($=\pm 1$) in Eq.~(\ref{EQ:UV-I}) and
(\ref{EQ:UV-II}), which will crucially influence the Josephson
current between two superconductors of dissimilar spin parity. We
shall also note that the electron component of the eigenfunction
is an even function of $\mathbf{k}$ (i.e.,
$u_{-\mathbf{k}}=u_{\mathbf{k}}$ arising from
$\varphi_{-\mathbf{k}}=\varphi_{\mathbf{k}}$) for the spin-singlet
pairing state while is an odd function of $\mathbf{k}$ (i.e.,
$u_{I(II),-\mathbf{k}}=-u_{I(II),\mathbf{k}}$ arising from
$\varphi_{-\mathbf{k}}=\varphi_{\mathbf{k}}+\pi$) for the triplet
pairing state.

The Josephson current $I_{J}$ originates from a
number of terms in the perturbation expression
Eq.~(\ref{EQ:JOSEPHSON}) in which the expectation value of two
creation operators in one superconductor is combined with the
expectation value of two annihilation operators in the other
superconductor, that is, $\langle
\tilde{c}_{\mathbf{k}\sigma_{1}}^{\dagger}(t)
\tilde{c}_{-\mathbf{k}\sigma_{2}}^{\dagger} (t^{\prime}) \rangle
\langle \tilde{c}_{\mathbf{p}\sigma_{1}^{\prime}}(t)
\tilde{c}_{-\mathbf{p}\sigma_{2}^{\prime}} (t^{\prime}) \rangle $.
Using the above symmetry properties, one can find the expectation
values for superconductors with a spin-singlet,
non-equal-spin-triplet, and equal-spin-triplet pairing state:
\begin{subequations}
\begin{eqnarray} & \langle
\tilde{c}_{\mathbf{k}\sigma}^{\dagger}(t)
\tilde{c}_{-\mathbf{k}\sigma^{\prime}}^{\dagger}(t^{\prime})
\rangle = \left( \begin{array}{c} \sigma
\delta_{\sigma,-\sigma^{\prime}}u_{\mathbf{k}}^{*}v_{\mathbf{k}}
\\ \delta_{\sigma,-\sigma^{\prime}}u_{I,\mathbf{k}}^{*}v_{I,\mathbf{k}}
\\ -\delta_{\sigma,\sigma^{\prime}}
u_{II,\mathbf{k}}^{*}v_{II,\mathbf{k}}
\end{array} \right)
&
\nonumber \\
&\times [e^{iE_{\mathbf{k}}(t-t^{\prime})}f(E_{\mathbf{k}})-
e^{-iE_{\mathbf{k}}(t-t^{\prime})}f(-E_{\mathbf{k}})]\;,&
\end{eqnarray}
and
\begin{eqnarray} & \langle
\tilde{c}_{\mathbf{p}\sigma}(t)
\tilde{c}_{-\mathbf{p}\sigma^{\prime}}(t^{\prime}) \rangle =
\left( \begin{array}{c} \sigma
\delta_{\sigma,-\sigma^{\prime}}u_{\mathbf{p}}v_{\mathbf{p}}^{*}
\\
\delta_{\sigma,-\sigma^{\prime}}u_{I,\mathbf{p}}v_{I,\mathbf{p}}^{*}
\\ -\delta_{\sigma,\sigma^{\prime}}
u_{II,\mathbf{p}}v_{II,\mathbf{p}}^{*}
\end{array} \right)
&
\nonumber \\
&\times [e^{-iE_{\mathbf{p}}(t-t^{\prime})}f(-E_{\mathbf{p}})-
e^{iE_{\mathbf{p}}(t-t^{\prime})}f(E_{\mathbf{p}})]\;,&
\end{eqnarray}
\end{subequations}
where the Fermi distribution function $f(E)=1/[\exp(E/T)+1]$. We
evaluate the Josephson current in various types of superconducting
junctions. First let us consider that both the left and right
superconductors are of spin-singlet pairing symmetry, one can
arrive at the Josephson current as:
\begin{eqnarray}
I_{J}&=&e\sum_{\mathbf{k},\mathbf{p}}
\sum_{\sigma\sigma^{\prime}}(\sigma\sigma^{\prime}) \mbox{Im} [
T_{\sigma\sigma^{\prime}}(t)T_{-\sigma,-\sigma^{\prime}}(t)
e^{i(2eVt+\delta \varphi)}] \nonumber \\
&& \times \frac{\vert\psi_{\mathbf{k}}\vert \vert
\psi_{\mathbf{p}}\vert
\Omega_{\mathbf{k},\mathbf{p}}(eV)}{2E_{\mathbf{k}}E_{\mathbf{p}}}\;,
\end{eqnarray}
where the phase difference
$\delta\varphi=(\varphi_{R}-\varphi_{L})+(\varphi_{\mathbf{p}}
-\varphi_{\mathbf{k}})$, and
\begin{eqnarray}
\Omega_{\mathbf{k},\mathbf{p}}(eV)&=&
[\frac{1}{eV+E_{\mathbf{k}}-E_{\mathbf{p}}}-
\frac{1}{eV-E_{\mathbf{k}}+E_{\mathbf{p}}}]
\nonumber \\
&& \times [f(E_{\mathbf{k}})-f(E_{\mathbf{p}})]\nonumber \\
&& +[\frac{1}{eV+E_{\mathbf{k}}+E_{\mathbf{p}}}-
\frac{1}{eV-E_{\mathbf{k}}-E_{\mathbf{p}}}] \nonumber \\
&& \times [1-f(E_{\mathbf{k}})-f(E_{\mathbf{p}})]\;.
\end{eqnarray}
The summation over spin indices involves the term,
$T_{\uparrow\uparrow}T_{\downarrow\downarrow}$, and
$T_{\uparrow\downarrow}T_{\downarrow\uparrow}$. It then follows
from the structure of the tunneling matrix as given by
Eq.~(\ref{EQ:SPIN}), which has the property
$T_{\downarrow\uparrow}=T_{\uparrow\downarrow}^{*}$, that the
flowing Josephson current is not modulated with time by the
precessing spin. Similarly, one can find that this conclusion is
also true for the Josephson current between two superconductors
both of non-equal-spin-triplet pairing symmetry. However, when
each side of the junction is a superconductor having
equal-spin-triplet pairing symmetry, the Josephson current
becomes:
\begin{eqnarray}
I_{J}&=&-e\sum_{\mathbf{k},\mathbf{p}}
\sum_{\sigma\sigma^{\prime}}
\mbox{Im}[T_{\sigma\sigma^{\prime}}(t)T_{\sigma\sigma^{\prime}}(t)
e^{i(2eVt+\delta \varphi)}] \nonumber \\
&& \times \frac{\vert d_{II,\mathbf{k}}\vert \vert
d_{II,\mathbf{p}}\vert
\Omega_{\mathbf{k},\mathbf{p}}(eV)}{2E_{\mathbf{k}}E_{\mathbf{p}}}\;,
\end{eqnarray}
which will be time dependent even in the absence of the voltage
bias when the spin is precessing at $\omega_{L}$. In some detail,
because $T_{\uparrow\downarrow}=T_{\downarrow\uparrow}^{*}=\vert
T_{\uparrow\downarrow}\vert e^{i\omega_{L} t}$, $I_{J}$ contains a
term with a pre-factor $\cos(2\omega_{L}t)$. This implies that the
Josephson current flowing between two equal-spin-triplet pairing
superconductors is modulated in time at a frequency of
$2\omega_{L}$, i.e., {\em twice of the Larmor frequency}. 
The relative ratio between the Larmor modulation part $\delta
I_{J}$ and the constant part $I_{J0}$ is:
\begin{equation}
\frac{\delta I_{J}}{I_{J0}}=\frac{J^{2}S^{2}}{2\Phi \Phi_{0}}\sim 10^{-2}
\mbox{--} 10^{-3}\;,
\end{equation}
for $\Phi=1\;\mbox{eV}$, $\Phi_{0}=0.05\;\mbox{eV}$, $JS=0.1\;\mbox{eV}$, 
which is experimentally detectable. The modulation of a Josephson current
by a precessing spin could be used for a single spin detection.

If we suppose that the left superconductor is a spin-singlet
superconductor which is weakly coupled to the right superconductor
having non-equal-spin-triplet pairing symmetry, the Josephson
current is found to be:
\begin{eqnarray}
I_{J}&=&e\sum_{\mathbf{k},\mathbf{p}}
\sum_{\sigma\sigma^{\prime}}\sigma \mbox{Im} [
T_{\sigma\sigma^{\prime}}(t)T_{-\sigma,-\sigma^{\prime}}(t)
e^{i(2eVt+\delta \varphi)}] \nonumber \\
&& \times \frac{\vert\psi_{\mathbf{k}}\vert \vert
d_{I,\mathbf{p}}\vert
\Omega_{\mathbf{k},\mathbf{p}}(eV)}{2E_{\mathbf{k}}E_{\mathbf{p}}}\;.
\end{eqnarray}
Notice that the summation $\sum_{\sigma\sigma^{\prime}} \sigma
T_{\sigma\sigma^{\prime}}(t)T_{-\sigma,-\sigma^{\prime}}(t)=
-\sum_{\sigma\sigma^{\prime}} \sigma
T_{\sigma\sigma^{\prime}}(t)T_{-\sigma,-\sigma^{\prime}}(t)$. This
property mandates that $I_{J}$ has to be zero. Also the Josephson
current cannot occur when a spin-singlet superconductor is weakly
coupled to a superconductor of equal-spin-triplet pairing
symmetry, due to the summation $\sum_{\sigma\sigma^{\prime}}\sigma
T_{\sigma\sigma^{\prime}}(t)T_{-\sigma,\sigma^{\prime}}(t)=0$.
Therefore, we conclude that the Josephson current cannot flow
between two superconductors with the pairing symmetry of different
spin parity even if there is a precessing spin located in the
tunnel barrier.

Based on the above microscopic analysis, we can establish a simple 
phenomenological theory for the Josephson effect through the
precessing spin. We define $\Psi_{\sigma}=\sigma \langle
c_{\mathbf{k}\sigma}c_{-\mathbf{k},-\sigma}\rangle$, $\langle
c_{\mathbf{k}\sigma}c_{-\mathbf{k},-\sigma}\rangle$, $\langle
c_{\mathbf{k}\sigma}c_{-\mathbf{k},\sigma}\rangle$ as the
macroscopic wave function on each side of the junction with
spin-singlet, non-equal-spin and equal-spin triplet, pairing
symmetry. The equations of motion can be written as:
\begin{subequations}
\begin{eqnarray}
i\frac{\partial\Psi_{L,\sigma}}{\partial t}&=&eV\Psi_{L,\sigma}
+\sum_{\sigma^{\prime}}
K_{\sigma\sigma^{\prime}}\Psi_{R,\sigma^{\prime}}\;,\\
i\frac{\partial\Psi_{R,\sigma}}{\partial t}&=&-eV\Psi_{R,\sigma}
+\sum_{\sigma^{\prime}}
K_{\sigma\sigma^{\prime}}^{*}\Psi_{L,\sigma^{\prime}}\;,
\end{eqnarray}
\end{subequations}
where $K_{\sigma\sigma^{\prime}}$ represents the spin-dependent
coupling across the barrier. Making the substitutions
$\Psi_{L(R),\sigma}=\mathcal{N}_{L(R),\sigma}^{1/2}
\exp(i\theta_{L(R)})$ with
$\mathcal{N}_{L(R)}=\sum_{\sigma}\mathcal{N}_{L(R),\sigma}$ the
number of Cooper pairs on each side, one can get the Josephson
current: $I_{J}=-\mathcal{N}_{L}^{1/2}\mathcal{N}_{R}^{1/2}
\mbox{Im}[K_{\sigma\sigma^{\prime}}\exp(i\delta\theta)]$ where
$\delta\theta=\theta_{R}-\theta_{L}=2eVt+\varphi_{R}-\varphi_{L}$.
For the junction formed by two spin-singlet paring
superconductors, all $K_{\sigma\sigma^{\prime}}$ are time
independent. For the spin-singlet/spin-triplet junction, the terms
contributing to the summation over spin indices cancel each
other. However, for the spin-triplet/spin-triplet junction, there
are terms proportional to $\vert K_{\uparrow\downarrow} \vert
\exp(\pm 2i\omega_{L}t)$. With this characteristics of $K$, one
can then arrive at the same conclusion as from the microscopic
analysis. We should stress that in the presence of spin-orbit
coupling, one is allowed to have direct coupling of a current
produced by spin-singlet superconductors and local spin
$\mathbf{S}$. This would lead to the time dependent contribution
of a Josephson current regardless of the pairing symmetry.

In summary, we have studied the Josephson current through a
precessing spin between various types of superconducting
junctions. It is shown that the Josephson current flowing between
two spin-singlet pairing superconductors is not modulated by the
precession of the spin. When both superconductors have
equal-spin-triplet pairing state, the flowing Josephson current is
modulated with twice of the Larmor frequency by the precessing
spin. It was also found that up to the second order in the
tunneling matrix elements no Josephson current can occur by the
direct exchange interaction between the localized spin and the
conduction electrons, if the two superconductors have different
spin-parity pairing states. As far as we know, no measurements of
Josephson current through a precessing spin between two
superconductors have been reported yet.  We feel the observation
of our predictions is within the reach of present technology. 
As a possible experiment, we mention results on  atomically sharp
superconducting tip in low temperature STM in both the
quasiparticle tunneling regime~\cite{Pan98} and the Josephson
tunneling regime~\cite{Naaman01} (coined as ``Josephson STM'' or
JSTM~\cite{Smakov01}) on conventional superconductors.
The spin-relaxation time in superconductor is
governed by the coupling of spin with its environment and is
strongly suppressed due to the gapped nature of quasiparticles, we
can expect a very long spin-relaxation time  when the local spin
is embedded in a superconducting junction. Therefore, it would be very 
interesting to extend the JSTM technology by using a superconducting
tip to study the Josephson current in the vicinity of an atomic spin
on the superconducting surface.

{\bf Acknowledgments}: We thank D. P. Arovas and M. Sigrist for
helpful discussions. This work was supported by the Department of
Energy.


\begin{thebibliography}{99}
\bibitem{Comment1} The feedback effects of a tunneling current on
a spin dynamics, a so-called back action effect, have been
considered in \cite{Mozy01,Balatsky02a,Balatsky02b}. It has been
shown that the back-action effects are small and precession is not
drastically affected by a tunneling current, provided the coupling is
small enough. This allows us to treat the problem with a long decay
time $\tau$ of precessing state $\omega_L \tau \gg 1$.
\end{thebibliography}
\end{document}